\documentclass[twocolumn,showpacs,preprintnumbers,nofootinbib,amsmath,amssymb,floatfix,aps,prc]{revtex4-1}
\usepackage{graphicx}% Include figure files
\usepackage{subfigure}
\usepackage{isotope}

\newcommand{\ve}[1]{\ensuremath{\mathbf{#1}}}
\newcommand{\n}[1]{\ensuremath{|\mathbf{#1}|}}

\newcommand{\Ek}{\ensuremath{E_k}}
\newcommand{\Ekp}{\ensuremath{E_{k'}}}
\newcommand{\Ep}{\ensuremath{E_p}}
\newcommand{\Epp}{\ensuremath{E_{p'}}}

\newcommand{\qrec}{\ensuremath{Q_{\textrm{rec}}^2}}
\newcommand{\qNC}{\ensuremath{Q_{\textrm{rec,\,NC}}^2}}
\newcommand{\qCC}{\ensuremath{Q_{\textrm{rec,\,CC}}^2}}
\newcommand{\erec}{\ensuremath{E_\nu^{\textrm{rec}}}}
\newcommand{\eB}{\ensuremath{\varepsilon}}
\newcommand{\sNC}{\ensuremath{\sigma^{\textrm{NC}}}}
\newcommand{\sCC}{\ensuremath{\sigma^{\textrm{CC}}}}
\newcommand{\MB}{MiniBooNE}
\newcommand{\MBC}{MiniBooNE Collaboration}
\def\lsim{\lesssim}

\def\be{\begin{equation}}
\def\ee{\end{equation}}

\begin{document}

\title{Consistent analysis of neutral- and charged-current neutrino scattering off carbon}
\author{Artur M. Ankowski}
\email{Artur.Ankowski@roma1.infn.it}
\affiliation{INFN and Department of Physics,``Sapienza'' Universit\`a di Roma, I-00185 Roma, Italy}

\date{\today}%

\begin{abstract}
\begin{description}
\item[Background] Good understanding of the cross sections for (anti)neutrino scattering off nuclear targets in the few-GeV energy region is a~prerequisite for the correct interpretation of results of ongoing and planned oscillation experiments.
\item[Purpose] Clarify a~possible source of disagreement between recent measurements of the cross sections on carbon.
\item[Method] Nuclear effects in (anti)neutrino scattering off carbon nucleus are described using the spectral function approach.
 The effect of two- and multinucleon final states is accounted for by applying an effective value of the axial mass, fixed to 1.23~GeV. Neutral-current elastic (NCE) and charged-current quasielastic (CCQE) processes are treated on equal footing.
\item[Results] The differential and total cross sections for the energy ranging from a~few hundreds of MeV to 100~GeV are obtained and compared to the available data from the BNL E734, MiniBooNE, and NOMAD experiments.
\item[Conclusions] Nuclear effects in NCE and CCQE scattering seem to be very similar. Within the spectral function approach, the axial mass from the shape analysis of the MiniBooNE data is in good agreement with the results reported by the BNL E734 and NOMAD Collaborations. However, the combined analysis of the NCE and CCQE data does not seem to support the contribution of multinucleon final states being large enough to explain the normalization of the MiniBooNE-reported cross sections.
\end{description}
\end{abstract}

\pacs{13.15.+g, 25.30.Pt}%
%13. Specific reactions and phenomenology
%13.15.+g 	Neutrino interactions
%25. Nuclear reactions: specific reactions
%25.30.Pt 	Neutrino-induced reactions

%\keywords{Suggested keywords}%Use showkeys class option if keyword display desired

\maketitle
\section{Introduction}
We are currently witnessing a~very rapid progress in the determination of neutrino oscillation parameters. The mixing angles and the mass-squared differences governing oscillations of solar~\cite{ref:SuperK,ref:Borexino,ref:KamLAND} and atmospheric~\cite{ref:MINOS_atm,ref:SuperK_atm} neutrinos are known with systematically improving precision; for a recent review see Ref.~\cite{ref:Schwetz}.

The past-year results from
the T2K~\cite{ref:T2K}, MINOS~\cite{ref:MINOS}, and Double Chooz~\cite{ref:DoubleChooz} experiments have brought indications that, in the three-neutrino framework, the last unknown mixing angle is nonvanishing. These hints have been confirmed by the findings from Daya Bay~\cite{ref:DayaBay}, followed by those from RENO~\cite{ref:RENO}, with statistical significance of 5.2 and 4.9 standard deviations, respectively.

Three nonvanishing mixing angles are a~prerequisite for the existence of the phase violating the charge-parity (CP) symmetry. Therefore, the results reported by the Daya Bay and RENO Collaborations are paving the way for studies of CP violation in the lepton sector, which, in turn, may bring us closer to understanding the matter-antimatter asymmetry in the universe. In addition, near-future experiments, such as NO$\nu$A, will aim at the determination of the mass hierarchy~\cite{ref:NOvA}.

The ongoing experiments might also be able to verify observations which do not seem to fit into the three-neutrino framework~\cite{ref:unexplained_LSND,ref:unexplained_MiniB}. If these revelations were confirmed, their explanation would require invoking sterile neutrinos~\cite{ref:sterile_Kopp,ref:sterile_Giunti,ref:sterile_Dib} and extending the standard model.

The correct interpretation of the outcome of oscillation experiments requires a~precise knowledge of the (anti)neu{\-}trino cross sections. This is the case even in those experiments in which the event yields in near and far detectors are compared, because, in general, flux and backgrounds do not scale in a~simple manner~\cite{ref:MiniB_CC}. If the cross section used in the analysis is underestimated in some energy region, a~false oscillation signal will be obtained as a~consequence.

The description of nuclear effects in neutrino scattering is now generally regarded as one of the main sources of systematic uncertainties in oscillation experiments. In particular, profound understanding of charged-current quasielastic (CCQE) interaction with nucleons bound in the nucleus plays a~crucial role. In the energy region of $\sim$1~GeV, it is the dominant reaction mechanism. For the energy of a~few GeV, $\Delta$-resonance excitation becomes equally important~\cite{ref:delta}. However, its description involves all---but not only~\cite{ref:Davide_delta}---the difficulties appearing in the CCQE case.

The carbon nucleus is of special importance, because the targets applied in neutrino detectors often involve carbon compounds, such as mineral oil, poly{\-}styrene, or organic scintillators. In addition, the proper understanding of carbon may provide useful clues for the description of oxygen, as nuclear effects in these two targets are expected to be similar, based on the experience gained in electron scattering.

The total $\isotope[12][6]{C}(\nu_\mu,\mu^-)$ cross section as a~function of energy has recently been measured with the NOMAD~\cite{ref:NOMAD} and \MB{}~\cite{ref:MiniB_CC} experiments.

In NOMAD, studying neutrinos of energy down to 3~GeV, CCQE events with and without knocked-out proton detected have been analyzed separately and, after adjusting the description of final-state interactions, shown to yield consistent results. Thanks to the 45-GeV average energy of the deep-inelastic-scattering events, the normalization has been determined from the well-known total inclusive charged-current (CC) cross section and from the purely leptonic process of inverse muon decay. The observed reduction of the cross section due to nuclear effects is $\sim$4\%.

In \MB{}, all events without pions detected have been classified as CCQE and the total cross section is extracted for energy up to 2~GeV. The high statistics of CCQE events ($\sim$$10\times$ those in NOMAD) allowed for obtaining, for the first time, the double differential cross section. Surprisingly, the reported total cross section for carbon is \emph{higher} than that for free neutrons. The size of the effect is $\sim$5\% at a~few hundred MeV, increasing to $\sim$15\% at neutrino energy higher than $\sim$850~MeV.

As a~20\% uncertainty of the cross section would have an important impact on determination of oscillation parameters~\cite{ref:Davide_PLB}, the difference between the NOMAD and \MB{} results requires very careful theoretical analysis, in addition to the planned investigation with the SciNO$\nu$A experiment~\cite{ref:SciNOvA}.

The \MBC{} has also measured the cross section for neutral-current elastic (NCE) interaction, in which nuclear effects are expected not to differ from those in the CCQE channel. The NCE data may thus provide a~cross-check for the applied description of the target nucleus and, what is even more important, additional information on nucleon kinematics in the \MB{} experiment, unavailable in CCQE scattering.
Therefore, in this paper, we pay special attention to NCE scattering.

A~complete description of a~nucleus should account for both the shell structure and nucleon-nucleon correlations of various kind. The shell structure is particularly important for studies involving low-energy probes. Correlations, causing strength fragmentation and partial depletion of the shells, modify the nuclear structure and give rise to deeply bound nucleons with high momentum, which are only accessible to high-energy probes.

Owing to the complicated nature of nuclear interactions and to the large number of nucleons, in the description of the carbon nucleus, approximations are necessary. One needs, however, to be aware of the accuracy of the approximations applied. For example, the nuclear transparency of carbon calculated without accounting for correlations is underestimated by $\sim$15\%~\cite{ref:Omar_transparency,ref:Daniela}.

The relativistic Fermi gas (RFG) model neglects both the shell structure and correlations, treating the nucleus as a~fragment of noninteracting nuclear matter of uniform density in a~constant potential. Due to its simplicity, it is applied in Monte Carlo generators used in data analysis. One needs to bear in mind, however, that obtaining good agreement between the simulation and the distributions of collected events may require introducing various {\it ad hoc} modifications, as a remedy for the shortcomings of the RFG approach~\cite{ref:NOMAD,ref:MiniB_kappa,ref:MINOS_QE}. Obviously, all modifications applied to experimental analysis must be justified on physics grounds within a~consistent theoretical scheme.

The \MB{} data have been analyzed within various models. Martini {\it et al.}~\cite{ref:Martini_CC} have developed an approach based on the random-phase approximation, which subsequently has been improved by introducing relativistic corrections~\cite{ref:Martini_CC&NC}. Mainly due to sizable contributions of the two-particle--two-hole and three-particle--three-hole processes, the results of Ref.~\cite{ref:Martini_CC&NC} are in excellent agreement with the \MB{} data for NCE and CCQE scattering, including those for the double differential CCQE cross section. Analysis of data from other experiments in the approach of Ref.~\cite{ref:Martini_CC&NC} is not yet available. It is somewhat surprising that for CCQE antineutrino interactions, Martini {\it et al.} predict significantly lower contributions of multinucleon processes and in the energy region of $\sim$1~GeV, the total cross section being similar to the RFG one.

This effect is not confirmed by the results of Nieves {\it et al.}~\cite{ref:Nieves_PRC,ref:Nieves_PLB}, obtained using an effective interaction determined from data for photon, electron, and pion scattering off nuclei. Calculations of the double differential $\nu_\mu$ CCQE cross section have the authors of Ref.~\cite{ref:Nieves_PLB} to the conclusion that the multinucleon contribution may effectively be accounted for by increasing the value of the axial mass. It is of great importance, e.g., for Monte Carlo simulations, allowing for incorporation of complicated processes in a~simple manner. In addition, comparing their results to the cross section extracted from \MB{}, Nieves {\it et al.} have deduced that the neutrino flux in \MB{} seems to be underestimated by $\sim$9\%. As in the calculations of Ref.~\cite{ref:Nieves_PLB} the multinucleon strength is constrained by precise data, this result points to a~possible source of discrepancy between the cross sections from NOMAD and \MB{}.

Amaro {\it et al.}~\cite{ref:RMF} have observed that the relativistic mean-field approach provides a~very good description of the shape of the double differential cross section extracted by \MB{}, but fails to reproduce its normalization. The authors of Ref.~\cite{ref:RMF} have hypothesized that accounting for meson-exchange currents could reduce the discrepancy.

Using the spectral function of carbon of Ref.~\cite{ref:Omar_LDA}, applied also in this paper, Benhar {\it et al.}~\cite{ref:Omar_paradigm} have shown that the \MB{} data for the differential cross section, be it as a~function of muon kinetic energy or its production angle, may be described with excellent accuracy when the value of the axial mass is adjusted. However, it turned out that the total cross section's dependence on neutrino energy fails to reproduce the data. This problem has been interpreted in Ref.~\cite{ref:Omar_paradigm} as likely to be related to the flux-unfolding procedure, required to extract the total cross section.

In an effort to understand the \MB{} data, Meucci {\it et al.}~\cite{ref:RGF_CC,ref:RGF_NC} have shown that the relativistic Green's function model may provide a~very good description of the total CCQE cross section while the  (double) differential NCE (CCQE) cross section is in fairly good agreement with the experimental points. The authors of Refs.~\cite{ref:RGF_CC,ref:RGF_NC} have ascribed the enhancement of the cross sections to rescattering of the knocked-out nucleon, which may also produce $\Delta$ resonance, and multinucleon processes. Within the approach based on the relativistic distorted-wave impulse approximation, Meucci {\it et al.}~\cite{ref:RGF_NC} have obtained the results in somewhat worse agreement with the NCE data, but similar to those of Ref.~\cite{ref:Butkevich_NC}.

Recently, Lalakulich {\it et al.}~\cite{ref:Mosel} have investigated the role of multinucleon effects in CC neutrino interactions using the Giessen Boltzmann-Uehling-Uhlenbeck event generator. In their calculations, the two-nucleon strength is fit to the \MB{} data using well-motivated physically {\it ansatz}. Note that in Ref.~\cite{ref:Mosel} the energy spectra of the knockout nucleons are presented in detail.

This paper provides an extension of the approach of Ref.~\cite{ref:gamma}, where NCE scattering off oxygen has been considered in the context of neutrino-induced gamma-ray production, to the carbon target. Some of the results for CCQE $\nu_\mu$ interaction have been previously presented (with arbitrary normalization) in Ref.~\cite{ref:ABF10}, discussing various issues related to the $Q^2$ distribution of event yield. Here we cover both NCE and CCQE processes, treating them in an identical manner and comparing the obtained cross sections to those measured by the Brookhaven National Laboratory Experiment 734 (BNL E734)~\cite{ref:BNL_E734_NC}, \MB{}~\cite{ref:MiniB_NC,ref:MiniB_CC}, and NOMAD~\cite{ref:NOMAD}.

Our numerical results differ from those presented in Ref.~\cite{ref:Omar_NC}, owing to an error found in the code used to obtain the cross sections presented there.

We acknowledge that the energy dependence of the cross sections for neutrino scattering off carbon has been also reported by the KARMEN~\cite{ref:KARMEN} and LSND~\cite{ref:LSND} Collaborations. However, in the energy region of these experiments, $\sim$30--50 MeV and $\sim$120--220 MeV for $\nu_e$ and $\nu_\mu$ data, respectively, scattering off the nucleus is known to be dominated by transitions between discrete states below the nucleon emission threshold. As we focus on nucleon knockout to the continuum, the description of such a~process lies beyond the scope of this work.

In Sec.~\ref{sec:kinematics} we consider in the most general manner the kinematics of nucleon knockout from a~nuclear target, recalling the definitions of reconstructed $Q^2$ applied in experimental analysis of NCE and CCQE neutrino interactions. Our approach is detailed in Sec.~\ref{sec:approach}. In Secs.~\ref{sec:NC} and~\ref{sec:CC} our results for NCE and CCQE scattering, respectively, are compared to the available experimental data. In Sec.~\ref{sec:Discussion} we discuss how nuclear effects influence the determination of the axial mass, and consider possible sources of discrepancy between different experiments. Finally, Sec.~\ref{sec:Summary} briefly summarizes our findings and states the conclusions.

\section{Reconstruction of $Q^2$ in neutrino interactions}\label{sec:kinematics}
In a~nucleon knockout from a~nucleus at rest by a~neutrino of energy $\Ek$, the energy and momentum conservation read, in general,
\begin{eqnarray}
\Ek+M_A&=&\Ekp+E_{A-1}+\Epp,\label{eq:enCons}\\
\ve k&=&\ve{k'}-\ve{p}+\ve{p'}\label{eq:momCons},
\end{eqnarray}
respectively, where $\Epp=\sqrt{M^2+\ve{p'}^2}$ is the energy of the knocked-out nucleon of momentum $\ve {p'}$. In the case of NCE (CCQE) interaction $\Ekp$ and $\ve{k'}$ denote the energy and momentum of the final neutrino (charged lepton of mass $m'$). The energy of the residual nucleus $E_{A-1}$ may be cast in the form
\[
E_{A-1}=\sqrt{(M_A-M+E)^2+\ve p^2},
\]
with $M$ ($M_A$) being the nucleon (target nucleus) mass. The removal energy $E$ is defined as the excitation energy of the residual nucleus in its rest frame. The momentum of the residual nucleus is denoted as $-\ve p$, to make use of a~simple interpretation of $\ve p$, being the struck nucleon's momentum when a~neutrino scatters off a~single nucleon in the nucleus.

In theoretical neutrino physics, the differential cross sections with respect to four-momentum squared $Q^2$,
\begin{eqnarray}
Q^2&=&(\ve{p'}-\ve{p})^2-(\Epp+E_{A-1}-M_A)^2\label{eq:Q2hadr}\\
&=&(\ve{k}-\ve{k'})^2-(\Ek-\Ekp)^2\label{eq:Q2lept}.
\end{eqnarray}
are often considered. However, invariant $Q^2$ cannot be calculated from measurable quantities and therefore in data analysis reconstructed $Q^2$, defined as
\begin{equation}
\qNC=2M(\Epp-M)\label{eq:recQ2NC}
\end{equation}
and
\begin{equation}
\qCC=2\erec(\Ekp-\n{k'}\cos\theta)-m'^2\label{eq:recQ2CC}
\end{equation}
in NCE and CCQE interactions, respectively, is used instead~\cite{ref:NOMAD, ref:MiniB_CC, ref:MiniB_NC}. The reconstructed energy of the incoming neutrino $\erec$, appearing in Eq.~\eqref{eq:recQ2CC}, is given by
\begin{equation}\label{eq:recE}
\erec= \frac{2\Ekp\tilde M-(m'^2+\tilde M^2 - M^2)}{2 ( \tilde M - \Ekp + \n{k'}\cos\theta )},
\end{equation}
where $\tilde M=M -\eB$, with $\eB$ being the average value of the removal energy.

The value of the $Q^2$ reconstructed in NCE (CCQE) interactions corresponds to the invariant $Q^2$ for the scattering off a~free (bound) nucleon at rest, yielding the measured value of the knocked-out nucleon energy (charged-lepton energy and its production angle~$\theta$), compare Eqs.~\eqref{eq:recQ2NC} and~\eqref{eq:recQ2CC} with Eqs.~\eqref{eq:enCons}--\eqref{eq:Q2lept}.

Differences between invariant $Q^2$ and $\qCC$ are discussed in Ref.~\cite{ref:ABF10}. Note that as in Eq.~\eqref{eq:recE} the expression $(\Ekp-\n{k'}\cos\theta)$ may be greater than $\tilde M$, the reconstructed energy does not have to be positive, leading to the negative values of $\qCC$ for certain kinematics.

\section{Description of the approach}\label{sec:approach}
We assume that the process of neutrino-nucleus interaction involves a~single nucleon and the remaining $(A-1)$ nucleons act as a~spectator system. This scheme, called the impulse approximation (IA), is valid when the momentum transfer $\n q=|\ve k-\ve{k'}|$ is high enough for the probe to distinguish individual nucleons in the nucleus, as the probe's spacial resolution is $\sim1/\n q$ (see Ref.~\cite{ref:Omar_oxygen}).

In the IA regime, the neutrino-nucleus cross section is obtained convoluting the elementary neutrino-nucleon cross section with the hole and particle spectral functions (SFs) of the nucleus.

The hole spectral function $P_\textrm{hole}(\ve p, E)$ is the probability distribution of removing a~nucleon of momentum $\ve p$ and leaving the residual nucleus with excitation $E$ in its rest frame. The particle spectral function $P_\textrm{part}(\ve{p'},\mathcal T')$ describes the propagation of a~nucleon of momentum $\ve{p'}$ and kinetic energy $\mathcal T'$.

In the relativistic Fermi gas model, frequently used in Monte Carlo simulations, the hole and particle SFs read
\[\begin{split}
P^\textrm{FG}_\textrm{hole}(\ve p, E)&=\frac{3}{4\pi p_F^3}\:\theta\big(p_F-\n p\big)\,\delta(\Ep-\varepsilon-M+E),\\
P^\textrm{FG}_\textrm{part}(\ve{p'},\mathcal T')&=\Big[1-\theta\big(p_F-\n{p'}\big)\Big]\,\delta(\Epp-M-\mathcal T'),
\end{split}\]
where $\Ep=\sqrt{M^2+\ve p^2}$, the average removal energy is denoted by $\varepsilon$, and $p_F$ is the Fermi momentum.

Note that, in general, the proton and neutron SFs may differ. However, for symmetric nuclei ($N=Z$) it turns out that the proton and neutron momentum distributions are the same and the difference between the SFs amounts to a~shift of the removal-energy distributions,
\begin{equation}\begin{split}
P^n_\textrm{hole}(\ve p,E)&=P^{p}_\textrm{hole}(\ve p,E-\Delta),\\
P^n_\textrm{part}(\ve{p'},\mathcal T')&=P^{p}_\textrm{part}(\ve{ p'},\mathcal T'),
\end{split}\end{equation}
taking into account that neutrons are more deeply bound. In the case of the carbon nucleus $\Delta=2.76$~MeV~\cite{ref:isotopes}.

Realistic hole spectral functions for various nuclei have been obtained by the authors of Ref.~\cite{ref:Omar_LDA} in the local density approximation (LDA), relying on a~premise that short-range correlations between nucleons are unaffected by surface and shell effects. This assumption is supported by the observation that for nuclei with mass number $A\geq4$, the momentum distribution (per nucleon) is largely independent of $A$ for momentum higher than $\sim$300~MeV~\cite{ref:Omar_RMP}. In the LDA scheme, the shell structure of the nucleus, deduced from experimental $(e,e')$ data~\cite{ref:Saclay_C, ref:Dutta}, can consistently be combined with the correlation contribution obtained from theoretical calculations for uniform nuclear matter at different densities~\cite{ref:Omar_NM,ref:Omar_LDA}. The carbon SF of Ref.~\cite{ref:Omar_LDA} has been extensively used in the analysis of electron scattering data in various kinematical regimes. Moreover, it provides a~quantitative account of the nucleon momentum distributions extracted from $(e,e'p)$ data at large missing energy and momentum~\cite{ref:Daniela}.

In the IA approach, the total NCE cross section reads
\begin{equation}\label{eq:xsec_NC}
\frac{d\sigma_{\nu A}^{\textrm{NC}}}{dQ^2}= \sum_{N=p,\,n}\int d^3p\,
dEP^N_\textrm{hole}(\ve p,E)\frac{d\sigma_{\nu N}^{\textrm{NC}}}{d Q^2 },
\end{equation}
where $\sigma_{\nu N}^{\textrm{NC}}$ is the elementary NCE neutrino-nucleon cross section,
\begin{equation}\label{eq:elementaryNCCS}
\frac{d \sigma_{\nu N}^{\textrm{NC}}}{d Q^2}=\frac{G^2_F}{8 \pi\Ek^2}\int d\omega\:\delta(\omega+M_A-E_{A-1}-\Epp)
\frac{L_{\mu \nu} \widetilde W^{\mu \nu}}{\Ep\Epp},
\end{equation}
with the Fermi constant $G_F=1.16637\cdot10^{-5}$/GeV$^2$. Although the Pauli blocking factor is not written explicitly, we account for the particle SF using the LDA result of Ref.~\cite{ref:ABF10}.

The contraction of the leptonic and hadronic tensors, $L_{\mu \nu}$ and $\widetilde W^{\mu \nu}$, may be cast in the form
\begin{equation}\begin{split}\label{eq:LH}
L_{\mu\nu}\widetilde W^{\mu\nu}&=2\sum_{j=1}^{5}A_j\widetilde W_j,
\end{split}\end{equation}
where $A_j$ are kinematic coefficients equal to
\begin{equation}\begin{split}\label{eq:A}
A_1&=2M^2k\cdot k', \\
A_2&=2\widetilde p\cdot k\:\widetilde p\cdot k'-M^2k\cdot k',\\
A_3&=2(\widetilde p\cdot k'\:k\cdot \widetilde q-\widetilde p\cdot k\:k'\cdot \widetilde q), \\
A_4&=2k\cdot \widetilde q\:k'\cdot \widetilde q-k\cdot k'\:\widetilde q^2, \\
A_5&=2(\widetilde p\cdot k\:k'\cdot \widetilde q+\widetilde p\cdot k'\:k\cdot \widetilde q-k\cdot k'\:\widetilde p\cdot \widetilde q),
\end{split}\end{equation}
with $\widetilde p=(\Ep,\:\ve p)$ and $\widetilde q=(\Epp-\Ep,\:\ve{p'} - \ve{p})$. The structure functions $\widetilde W_j$ are related to the nucleon form factors by
\begin{equation}\begin{split}\label{eq:structureFunctions}
\widetilde W_1&=\tau\big(\mathcal{F}_1^N+\mathcal{F}_2^N\big)^2 + (1 +\tau)\big(\mathcal{F}_A^N\big)^2, \\
\widetilde W_2&=\big(\mathcal{F}_1^N\big)^2 + \tau\big(\mathcal{F}_2^N\big)^2+\big(\mathcal{F}_A^N\big)^2, \\
\widetilde W_3&=\big(\mathcal{F}_1^N+\mathcal{F}_2^N\big)\mathcal{F}_A^N, \\
\widetilde W_4&=\frac14\Big[\big(\mathcal{F}_1^N\big)^2 + \tau\big(\mathcal{F}_2^N\big)^2-\big(\mathcal{F}_1^N+\mathcal{F}_2^N\big)^2\\
&\quad-4\mathcal{F}_P^N\big(\mathcal{F}_A^N-\tau\mathcal{F}^N_P\big)\Big],\\
\widetilde W_5&=\frac{1}{2}W_2,
\end{split}\end{equation}
with $\tau=-\widetilde q^2/(4M^2)$. In the case of antineutrino scattering, the sign of $\widetilde W_3$ is reversed. The form factors appearing in NCE scattering ($\mathcal{F}_i^N$, $\mathcal{F}_A^N$, and $\mathcal{F}_P^N$) can be expressed by those known from the electromagnetic and CCQE interactions
($F^N_i$, $F_A$, and $F_P$) as
\begin{equation}\begin{split}
\mathcal{F}_i^N&=\pm \frac{1}{2}(F_i^p-F_i^n)-2\sin^2\theta_W F_i^N,\\
\mathcal{F}_A^N&=\frac{1}{2}\big(F_A^s\pm F_A\big)=\frac{1}{2}\frac{\Delta s\pm g_A}{(1-\widetilde q^2/M^2_A)^2},\\
\mathcal{F}_P^N&=\frac{2M^2\mathcal{F}_A^N}{m^2_\pi -\widetilde q^2}
\end{split}\end{equation}
(see Refs.~\cite{ref:Weinberg,ref:Donnelly&Peccei,ref:Alberico}). In the above equations, the upper (lower) sign refers to the proton (neutron) form factors, the applied value of the weak mixing angle $\theta_W$ is such that $\sin^2\theta_W=0.23116$, $m_\pi$ is the pion mass, $g_A=-1.2673$, and the strange quark contribution to the axial form factor is set to
$\Delta s=-0.08$~\cite{ref:strange}. As far as the strange contributions to $\mathcal{F}_i^N$ are concerned, in our approach they are neglected on the basis of experimental evidence~\cite{ref:G0,ref:HAPPEX}, on the one hand, and lack of reliable theoretical guidance on their $Q^2$ dependence, on the other hand.

The results presented in this article are obtained using state-of-the-art parametrization of the measurable Sachs form factors $G^N_E$ and $G^N_M$ from Refs.~\cite{ref:FormFactors_p,ref:FormFactors_n}, related to the electromagnetic form factors
$F_i^N$ by
\begin{equation}\begin{split}
F_1^N = \frac{G^N_E+\tau G^N_M}{1+\tau},\qquad F_2^N = \frac{G^N_M-G^N_E}{1+\tau}.
\end{split}\end{equation}

The dipole parametrization of the axial form factor is employed. Calculating the cross sections on free nucleons, we use the world average axial mass value of 1.03~GeV~\cite{ref:Meissner}, in good agreement with the recent analysis~\cite{ref:BBA03} of deuteron measurements~\cite{ref:BNL81,ref:ANL82,ref:BNL90}. In scattering on nucleons bound in the carbon nucleus, the effective $M_A=1.23$~GeV is applied, which has been determined by the \MBC{} from the first shape analysis of the $\qrec$ distribution of the largest statistics of CCQE events collected to date~\cite{ref:MiniB_kappa}. It seems to be well justified, as for given $M_A$ the approach of Ref.~\cite{ref:MiniB_kappa} and our calculation yield the shape of the \MB{} flux-averaged $\qrec$ distribution of CCQE events in very good agreement over a~broad range of $\qrec$~\cite{ref:ABF10}. While this purely phenomenological method of accounting for multinucleon processes in the carbon nucleus cannot be expected to be accurate in any kinematical setups, its validity is quantitatively supported by the results of Nieves {\it et al.}~\cite{ref:Nieves_PLB} for the double differential cross section in a~broad kinematical range of \MB{}. In addition, it predicts nuclear effects in CCQE $\nu_\mu$ and $\bar\nu_\mu$ scattering to be similar, consistently with the conclusion from the NOMAD experiment~\cite{ref:NOMAD}.
%consistently with the cross sections reported from the NOMAD experiment~\cite{ref:NOMAD}.

As neutrinos may undergo CCQE scattering only off neutrons, the cross section reads in this case
\begin{equation}\label{eq:xsec_CC}
\frac{d\sigma_{\nu A}^{\textrm{CC}}}{dQ^2}= \int d^3p\,
dEP^n_\textrm{hole}(\ve p,E)\frac{d\sigma^{\textrm{CC}}}{d Q^2},
\end{equation}
where the elementary CCQE cross section $\sigma^{\textrm{CC}}$ may be obtained from Eqs.~\eqref{eq:elementaryNCCS}--\eqref{eq:structureFunctions}, replacing
\begin{equation}\begin{split}
G_F&\rightarrow G_F\cos\theta_C,\\
\Ekp=\n{k'}&\rightarrow\Ekp=\sqrt{m'^2+\ve{k'}^2},\\
\mathcal{F}_i^N&\rightarrow F_i^p-F_i^n,\\
\mathcal{F}_A^N&\rightarrow F_A.\\
\end{split}\end{equation}
The value of the Cabibbo angle $\theta_C$ corresponds to $\cos\theta_C=0.97418$.

It is worth noting that the differences between the expressions for the NCE cross section given in this paper and those in Ref.~\cite{ref:Horowitz}, describing nuclear effects within the relativistic Fermi gas model, can be traced back to a~different treatment of the off-shell dynamics.

In our approach, off-shell effects are accounted for by using the de Forest approximation~\cite{ref:deForest}, which amounts to describing the process of scattering on an off-shell nucleon as if it involved a~free nucleon which absorbed only part of the energy transferred by the probe. Note that in Eq.~\eqref{eq:LH}, even the terms vanishing on the mass shell are kept. For example, the structure function $\widetilde W_4$ contributes to the NCE cross section only due to off-shell effects, because when $\Epp-\Ep=\Ek-\Ekp$ the coefficient $A_4$ is proportional to the neutrino mass. Neglecting $\widetilde W_4$ increases the total $\isotope[12][6]{C}(\nu,\nu')$ cross section by 1.4\% at $\Ek=450$~MeV. For higher and lower energy the effect, resulting mainly from the absence of $F_P$, is larger, reaching 3.0\% at 100 MeV and 2.0\% at 5 GeV.

The strength of final state interactions (FSI) in neutrino scattering can be deduced from comparison between electron scattering data and theoretical results, although the picture is obscured by flux averaging.

The effect of FSI is shown to be sizable in the relativistic and semirelativistic approaches of Refs.~\cite{ref:Caballero,ref:FSI,ref:Meucci}, which account for it by means of strong relativistic potentials or their semirelativistic equivalents. Yielding the overwhelming contribution to the high-$\omega$ tail of the $(e,e')$ cross section~\cite{ref:Caballero}, FSI play in these mean-field models an essential role in bringing the results in good agreement with data~\cite{ref:RMF}.

Within the SF approach, FSI are much weaker. The tail of the cross section originates from short-range correlations between nucleons in the initial state and is only slightly enhanced by FSI~\cite{ref:Omar_oxygen}. As observed in Ref.~\cite{ref:Omar_paradigm}, in most of the kinematical region spanned by the \MB{} data set, the main effect of FSI on the double differential $(e,e')$ cross section is a shift of the quasielastic peak of the order of 10~MeV. Using the energy-dependent and $A$-independent fit to the Dirac optical potential of carbon~\cite{ref:Cooper,ref:Hama}, we find the shift averaged over the \MB{} kinematics to be 6.8 MeV, consistent with the observation of the authors of Ref.~\cite{ref:Omar_paradigm}. At higher momentum transfers, allowed in NOMAD, FSI are known to lead to a redistribution of the inclusive strength, resulting in the quenching of the quasielastic peak and the enhancement of its tails~\cite{ref:Omar_RMP,ref:FSI}.

(Quasi)elastic processes constrain the high-$\n q$ contribution to appear solely at high $Q^2$, making it negligibly small due to the nucleon form factors. For example, while at a~neutrino energy of 100 GeV momentum transfers up to 100.85 GeV are allowed, as much as 89.8 (97.5)\% of the $\isotope[12][6]{C}(\nu_\mu,\mu^-)$ cross section comes from $\n q\leq1.645$ (2.747)~GeV, the range of $\n q$ for $\Ek=1$ (2)~GeV. Hence, in the context of our work, the kinematical setup of NOMAD does not differ significantly from that of \MB{}.

In this paper, we consider the total cross sections and the flux-averaged differential cross sections $d\sigma/dQ^2$ for neutrino and antineutrino (quasi)elastic scattering. As FSI may cause only a~redistribution and a~shift of the strength, they do not affect the total cross section. The differential cross sections $d\sigma/dQ^2$ are expected to be modified by FSI at $Q^2 \lsim 0.15$ GeV$^2$~\cite{ref:Omar_oxygen},\footnote{Note that this observation is largely independent of neutrino energy and therefore it applies also to flux-averaged results.} because only at this kinematics the real part of the optical potential significantly changes the typical energy of the knocked-out nucleon, compare to Eq.~\eqref{eq:recQ2NC}. However, in the low-$Q^2$ regime, the validity of the impulse approximation, underlying our calculations, becomes questionable~\cite{ref:ABF10,ref:IA}. Therefore, in this article, FSI are not taken into account.

\section{Comparison to the NCE data}\label{sec:NC}
From the neutrino experiments which obtained results on NCE scattering off nuclei, the highest event statistics to date have been collected by the BNL E734 and \MB{} Collaborations, allowing for extraction of the differential cross sections. Earlier experiments~\cite{ref:CIR,ref:MPW,ref:Gargamelle,ref:BMPW,ref:CERN,ref:CIB,ref:PHB}, aiming at verification of the theory of electroweak interactions, reported the total NCE to CCQE event numbers rates for neutrinos and antineutrinos.

In the BNL E734 experiment~\cite{ref:BNL_E734_NC}, studying $\nu p$ and $\bar\nu p$ NCE interactions, the target was composed in 79\% of protons bound in carbon and aluminium and in 21\% of free protons. The mean value of neutrino (antineutrino) energy was 1.3~GeV (1.2~GeV). Determination of the $\nu$ and $\bar\nu$ fluxes involved fitting to the CCQE event sample~\cite{ref:BNL_E734_flux}, under the assumption that the CCQE cross sections can be described as in the paper of Llewellyn-Smith~\cite{ref:Llewellyn-Smith} with corrections for Fermi motion and Pauli blocking~\cite{ref:Bell}. The dipole parametrization of the vector and axial form factors was applied, with the cutoff masses $M_V=0.84$~GeV and $M_A=1.03$~GeV, respectively.

Based on 1686 (1821) candidate $\nu p$ ($\bar\nu p$) events surviving the cuts, the flux-averaged differential cross sections ${d\sNC}/{dQ^2}$ for scattering off the BNL E734 target were extracted~\cite{ref:BNL_E734_NC}. The  normalization uncertainty of the $\nu$ ($\bar\nu$) result is estimated to be 11.2\% (10.4\%).

We calculate the cross sections for the BNL E734 target, neglecting the contribution of aluminium nuclei, according to
\be\label{eq:BNL}
\frac{d\sNC}{dQ^2}=f\frac{d\sigma_{\nu p,\,H}}{dQ^2}+(1-f)\frac{d\sigma_{\nu p ,\,C}}{dQ^2},
\ee
where $f=0.21$ is the fraction of free nucleons in the target, whereas $\sigma_{\nu p,\,H}$ ($\sigma_{\nu p ,\,C}$) denotes the cross section for scattering off free protons (protons bound in carbon) averaged over the neutrino energy $0.2\leq\Ek\leq5$~GeV.

\begin{figure}
\centering
    \includegraphics[width=0.80\columnwidth]{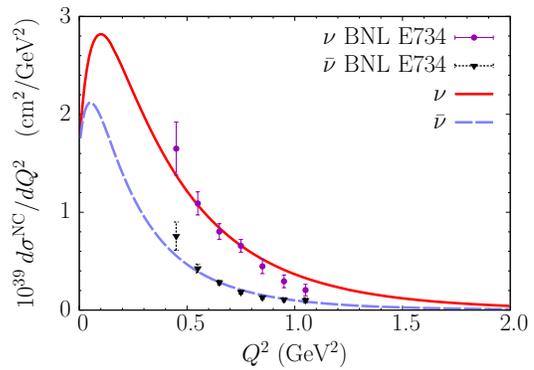}
\caption{\label{fig:BNL_Q2}(Color online) Differential cross section $\sNC/dQ^2$ for NCE $\nu$ and $\bar\nu$ scattering in the BNL E734 experiment. Our results are compared to the data from Ref.~\cite{ref:BNL_E734_NC}. The error bars do not include the normalization uncertainty of 11.2\% (10.4\%) in the $\nu$ ($\bar\nu$) case.
}
\end{figure}

Figure~\ref{fig:BNL_Q2} shows that the SF calculations provide a~fairly good description of the BNL E734 data~\cite{ref:BNL_E734_NC} in the whole range of $Q^2$. Note that the agreement seems to be better with the
higher-statistics antineutrino data. This is also the case for neutrinos in the region of the lowest uncertainty, $0.5\leq
Q^2\leq0.8$~GeV$^2$.

We checked that using $\qrec$ defined in Eq.~\eqref{eq:recQ2NC} instead of $Q^2$ does not change the conclusions significantly, as the main difference between these two variables appears for the values lower than 0.13~GeV$^2$.

The fluxes, and thus the absolute normalization of the BNL E734 data, were determined using a~description of CCQE interaction clearly different from ours. Do, therefore, our calculations agree with the data in just an accidental way? The flux-averaged $\isotope[12][6]{C}(\nu,\nu')$ cross section obtained with $M_A=1.23$ GeV varies by a~factor of 9.1 at $0.15\leq Q^2\leq 1.15$ GeV$^2$. However, over this interval, it differs by less than 10\% from the free cross section calculated with $M_A=1.03$ GeV. The corrections for Pauli blocking applied by the authors of Ref.~\cite{ref:BNL_E734_NC} could diminish this difference, which suggests that numerically our results may not be very different from those used to find the normalization of the BNL E734 cross sections.

We would like to acknowledge that the BNL E734 data have been analyzed by the authors of Refs.~\cite{ref:Garvey,ref:BNL_E734_reanalysis} to obtain constraints on the strange contribution to the form factors.

The \MB{} experiment, using the Cherenkov detector filled with mineral oil (CH$_2$), is sensitive to both $\nu p$ and $\nu n$ NCE scattering~\cite{ref:MiniB_NC}. In neutrino mode, its beam has been composed almost exclusively of muon neutrinos, with an average energy of 788~MeV. The neutrino flux at the detector has been determined by a~Monte Carlo simulation covered in detail in Ref.~\cite{ref:MiniB_flux}. The accuracy of the shape determination has been proven by the comparison of the observed
and predicted $\nu_\mu$ energy distribution in the CCQE event sample. The calculations were performed within the RFG model with enhanced Pauli blocking effect. The dipole parametrization of the axial form factor was used with $M_A=1.23$~GeV, as in this paper. However, the normalization of the measured distribution of CCQE events is reported to be higher by a~factor of $1.21\pm0.24$ than the calculated one~\cite{ref:MiniB_flux}.

The \MBC{} recorded in neutrino mode 94531 candidate events surviving the NCE cuts, which allowed for extraction of the differential cross section with unprecedented precision.

The theoretical estimate of the differential NCE cross section \emph{per nucleon} as a~function of $Q^2$ reads for the \MB{} target
\be\label{eq:dsNC/dQ2}
\frac{d\sNC}{dQ^2}=\frac{2}{14}\frac{d\sigma_{\nu p,\,H}}{dQ^2}+\frac{6}{14}\sum_{N=p,\,n}\frac{d\sigma_{\nu N,\,C}}{dQ^2}.
\ee
The most important difference with respect to Eq.~\eqref{eq:BNL} is the contribution of the neutrons bound in the carbon nucleus.

\begin{figure}
\centering
    \subfigure
    {\label{fig:NC_lin_recQ2}
    \includegraphics[width=0.80\columnwidth]{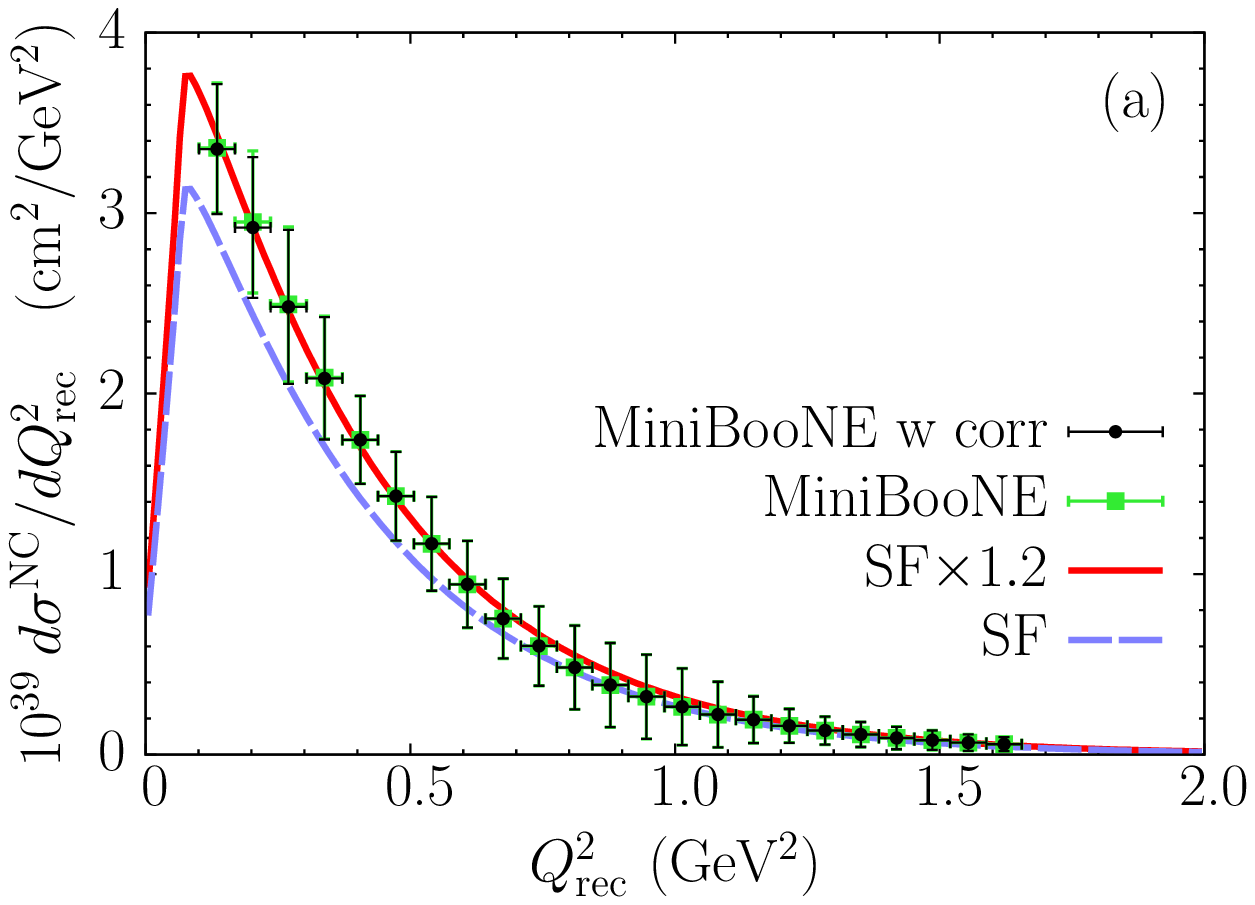}}
    \subfigure
    {\label{fig:NC_lin_Q2}
    \includegraphics[width=0.80\columnwidth]{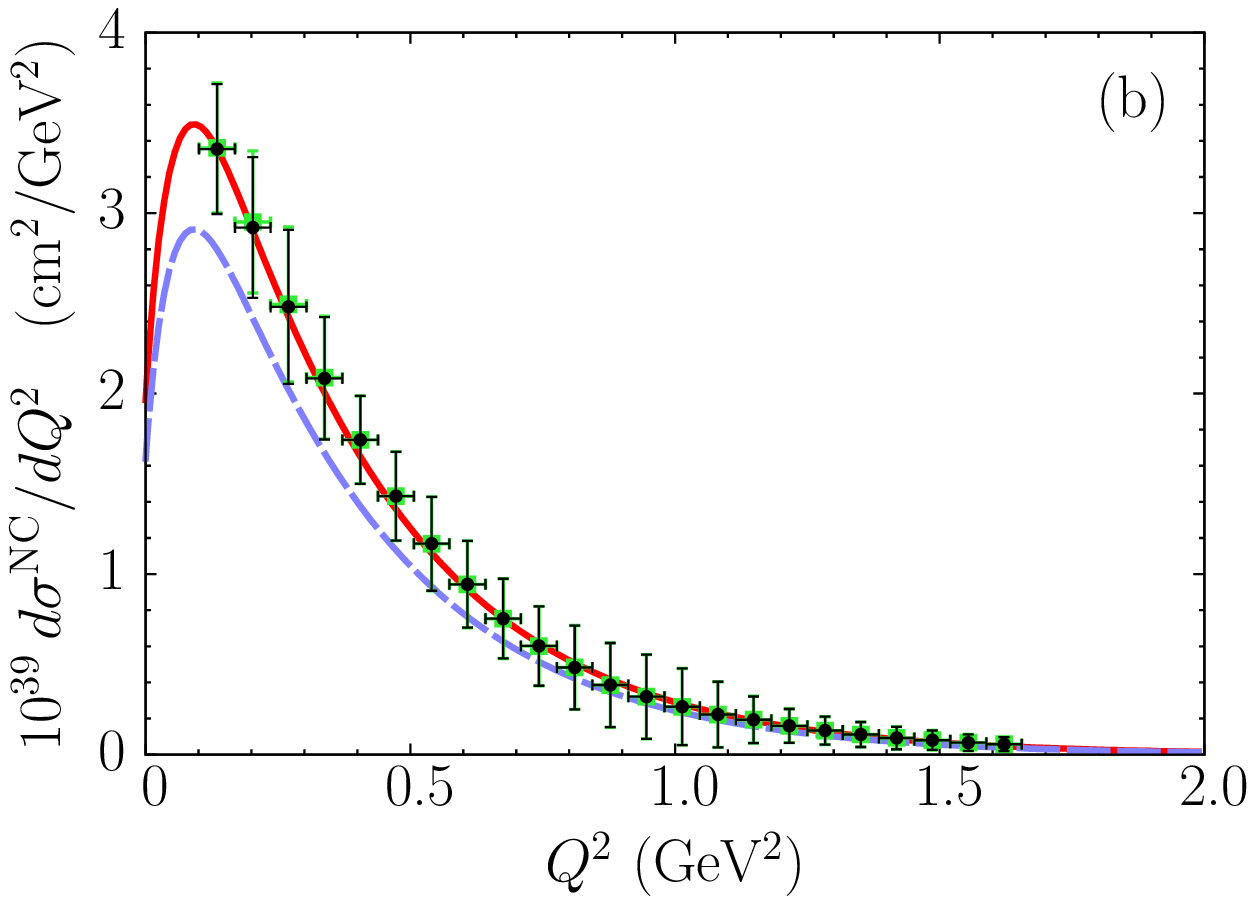}}
\caption{\label{fig:NC_lin}(Color online) Differential cross section $d\sNC/d\qrec$ [Panel (a)] and $d\sNC/dQ^2$ [Panel (b)] for NCE neutrino scattering off CH$_2$ averaged over the \MB{} flux. The spectral function calculations (dashed line) are compared to the data~\cite{ref:MiniB_NC} with (circles) and without (squares) the correction~\eqref{eq:corr}. The error bars do not account for the normalization uncertainty of 18.1\%. The absolute normalizations agree when our results are multiplied by 1.2 (solid line).
}
\end{figure}

\begin{figure}
\centering
    \subfigure
    {\label{fig:NC_log_recQ2}
    \includegraphics[width=0.80\columnwidth]{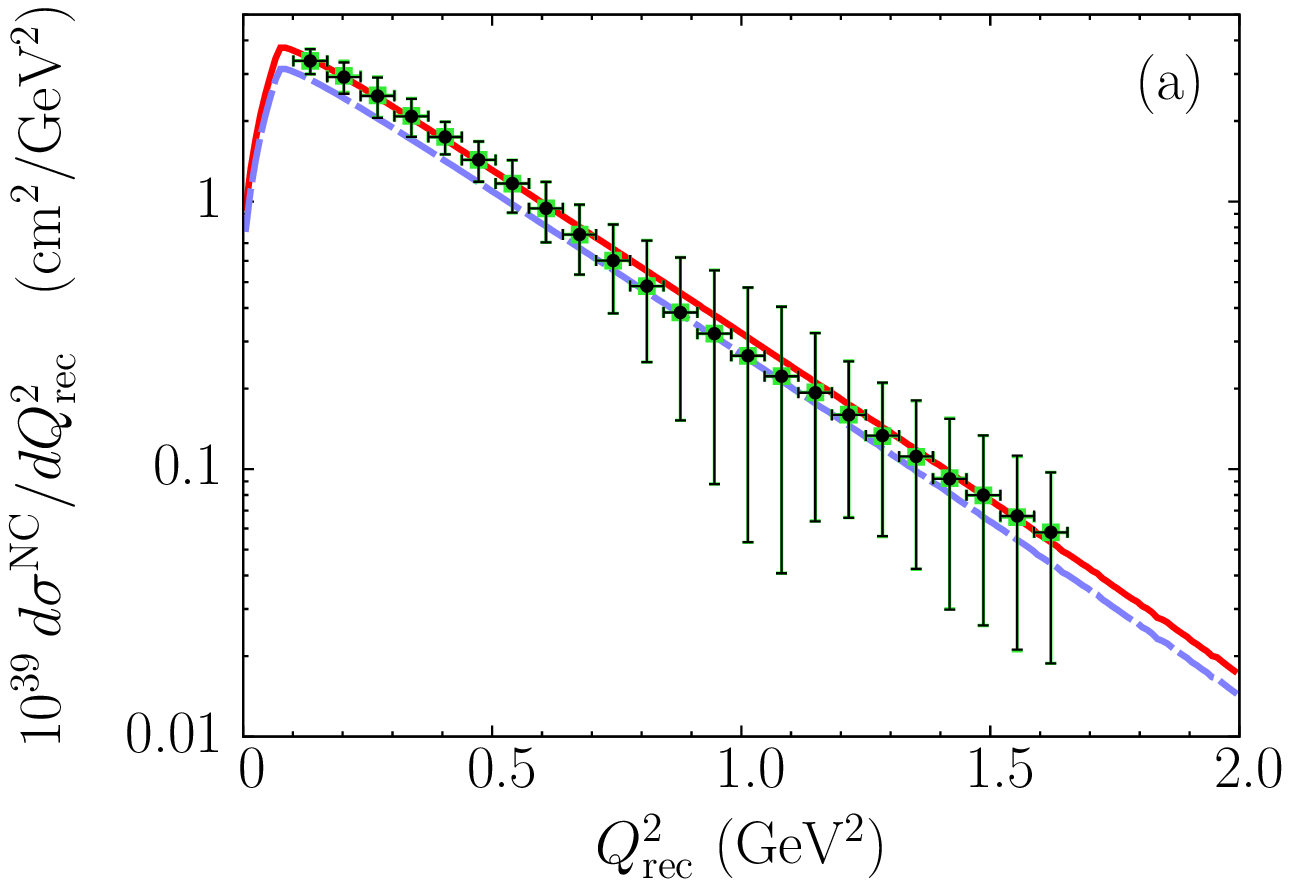}}
    \subfigure
    {\label{fig:NC_log_Q2}
    \includegraphics[width=0.80\columnwidth]{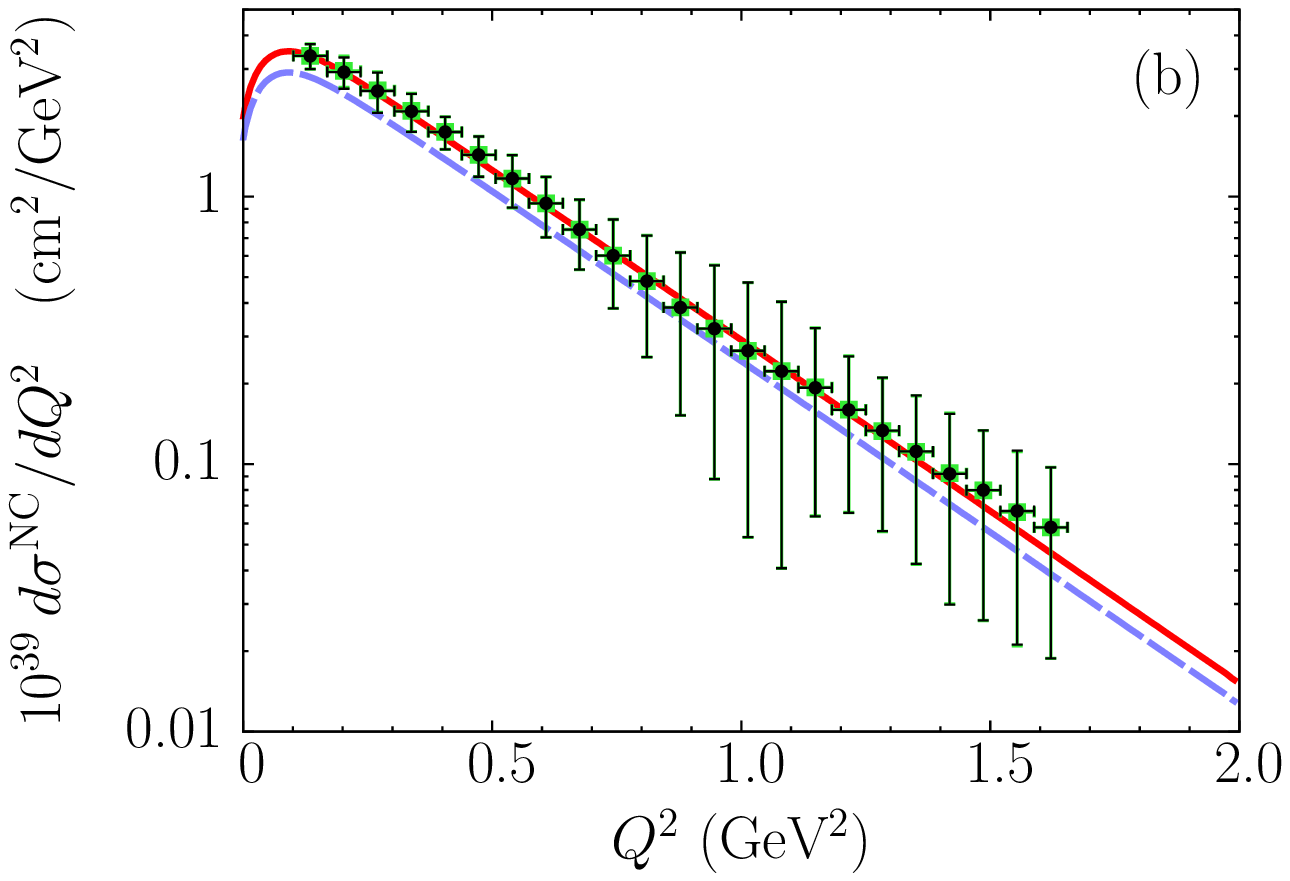}}
\caption{\label{fig:NC_log}(Color online) Same as Fig.~\ref{fig:NC_lin} but on a logarithmic scale.
}
\end{figure}

The detection efficiency for each contributing process is estimated in Monte Carlo simulations to be different; therefore the \MBC{} reported in Ref.~\cite{ref:MiniB_NC} both the cross section,
\be
\frac{d\sNC_\textrm{exp}}{d\qrec}=\frac{2}{14}C_{\nu p,\,H}\frac{d\sigma_{\nu p,\,H}}{dQ^2}+\frac{6}{14}\sum_{N=p,\,n}C_{\nu N,\,C}\frac{d\sigma_{\nu N,\,C}}{d\qrec},
\ee
and the dependence of the efficiency corrections $C_i$ on $\qrec$, being defined as in Eq.~\eqref{eq:recQ2NC}. Note that in scattering off a~free proton $\qrec=Q^2$.

Within a~given approach, it is possible to obtain a single effective correction to the theoretical calculations $C$, such that
\be\label{eq:C}
\frac{d\sNC_\textrm{exp}}{d\qrec}=C\frac{d\sNC}{d\qrec},
\ee
with ${d\sNC}/{d\qrec}$ being defined by analogy to Eq.~\eqref{eq:dsNC/dQ2}.

Because the values of the efficiency corrections are given only for the values of $\qrec$ corresponding to the reported cross section, it is convenient to apply the correction to the data rather than to the theoretical result, according to
\be\label{eq:corr}
\frac{1}{C}\frac{d\sNC_\textrm{exp}}{d\qrec}.
\ee

In general, such a~correction is a~model-dependent function of the axial mass. It turns out, however, that in our calculations, the effects of the proton and neutron corrections nearly cancel each other and the value of $C$ is very close to 1. The deviations of $C$ from the unit value, being typically less than 0.25\%, exceed 1\% in only one bin of $0.169\leq \qrec\leq0.236\textrm{ GeV}^2$. Therefore, the \MB{} data with and without the correction~\eqref{eq:corr} largely overlap.

Figures~\ref{fig:NC_lin} and~\ref{fig:NC_log} show that our calculations reproduce the \emph{shape} of the \MB{} NCE $\nu$ cross section very accurately, both in the region of lower $\qrec$ and in the tail. For $\qrec\leq0.64$~GeV$^2$, the discrepancies are on average~1.6\%, remaining smaller than $2.7\%$. The most sizable deviations appear for $0.8\leq\qrec\leq1.1$~GeV$^2$, where also the data uncertainties are the largest. However, to match the absolute scale of the \MB{} data, it was necessary to multiply the SF results by a~factor of 1.2. Although this factor is somewhat larger than the data normalization uncertainty of 18.1\% reported in Ref.~\cite{ref:MiniB_NC}, it remains consistent with the normalization discrepancy of $1.21\pm0.24$ observed by the \MBC{} in the CCQE analysis~\cite{ref:MiniB_kappa}.

Using in comparisons with the \MB{} NCE data the differential cross section as a~function of the theoretical $Q^2$ instead of $\qrec$, as in Refs.~\cite{ref:Omar_NC,ref:Martini_CC&NC,ref:Meucci_NC}, is a~fair approximation. The differences between them are easiest to notice for $\qrec\lesssim0.13\textrm{ GeV}^2$, compare panels (a) and (b) of Fig.~\ref{fig:NC_lin}, i.e., in the region where no experimental data are available. However, when the cross section's slope on a~logarithmic scale is of interest, the discrepancy, exceeding 10\%, appears also for $\qrec\gtrsim1\textrm{ GeV}^2$ (see Fig.~\ref{fig:NC_log}). We observe that for $0.12\leq\qrec\leq1.75\textrm{ GeV}^2$, the effect of using $\qrec$ may be approximated by
\be
\frac{d\sNC}{d\qrec}\approx(1+aQ^2)\frac{d\sNC}{dQ^2}
\ee
with 2\% accuracy. When the axial mass of 1.23~GeV is applied, the parameter $a$ is equal to 0.095/GeV$^2$ (0.070/GeV$^2$) for the SF approach (RFG model).

\begin{figure}
\centering
    \includegraphics[width=0.80\columnwidth]{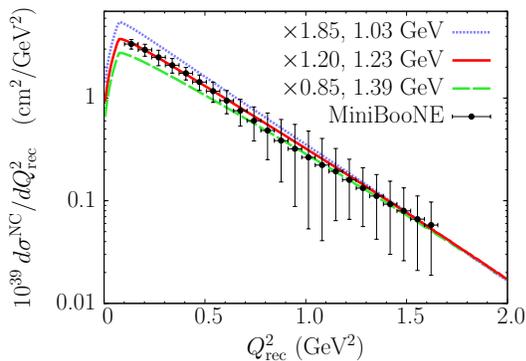}
\caption{\label{fig:MA}(Color online) Dependence of the shape of the differential NCE cross section ${d\sNC}/{d\qrec}$ in the \MB{} experiment on the axial mass value. The curves are labeled by $M_A$ and the multipliers applied to match their values at high $\qrec$.
}
\end{figure}

It is worth emphasizing that within the approach of this paper, it does not seem possible to achieve a~good agreement with the shape of the \MB{} NCE data applying an axial mass very different from 1.23~GeV. To illustrate this, in Fig.~\ref{fig:MA} we show the results obtained with $M_A$ equal to 1.03, 1.23, and 1.39~GeV. Only the middle value of the axial mass yields the correct slope of the NCE cross section.

To clarify a~possible source of the normalization discrepancy, it is useful to check how our approach describes the CCQE data.

\section{Comparison to the CCQE data}\label{sec:CC}
For CCQE neutrino scattering, the \MBC{} has obtained the differential cross section as a~function of $\qrec$, defined as in Eq.~\eqref{eq:recQ2CC}, from the collected in neutrino mode 146070 events passing the cuts~\cite{ref:MiniB_CC}. Note that the analysis in Ref.~\cite{ref:MiniB_CC} differs in the background treatment from the earlier one, described in Ref.~\cite{ref:MiniB_kappa} (193709 events passing the cuts).

In Ref.~\cite{ref:MiniB_CC} is also reported the double differential cross section $d\sCC/dT_\mu d\cos\theta$, with $T_\mu$ and $\cos\theta$ being the muon kinetic energy and production angle, respectively. As this paper aims at treating NCE and CCQE interaction on equal footing, we discuss only the data for the single differential cross section.

We have calculated the differential CCQE cross sections $d\sCC/dQ^2$ and $d\sCC/d\qrec$, as in Eq.~\eqref{eq:xsec_CC}, and averaged them over the \MB{} flux.

\begin{figure}
\centering
    \includegraphics[width=0.80\columnwidth]{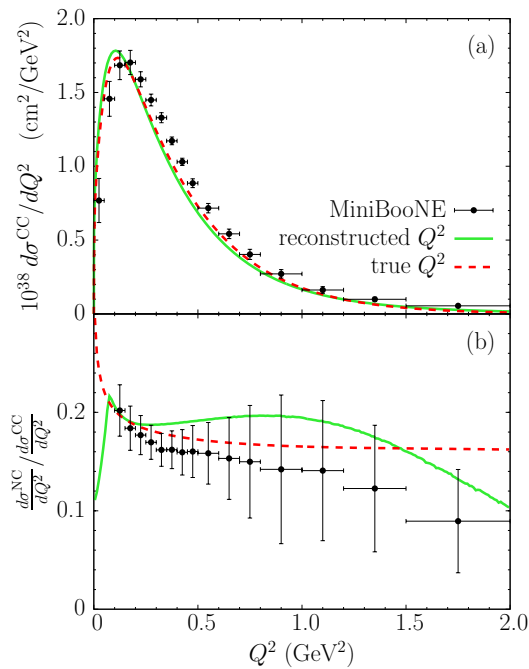}
    \subfigure{\label{fig:CC_a}}
    \subfigure{\label{fig:CC_b}}
\caption{\label{fig:CC}(Color online) (a) The \MB{} flux-averaged differential cross section for the $\isotope[12][6]{C}(\nu_\mu,\mu^-)$ scattering as a function of $\qrec$ (solid line) and $Q^2$ (dashed line). Our results multiplied by 1.2 are compared to the data~\cite{ref:MiniB_CC}. The error bars do not include the normalization uncertainty of 10.7\%. (b) The ratio of the NCE to CCQE cross sections per nucleon for CH$_2$ in the \MB{} experiment. The line code is the same as that for panel (a). The data are taken from Ref.~\cite{ref:MiniB_NC}.
}
\end{figure}

\begin{figure}
\centering
    \includegraphics[width=0.80\columnwidth]{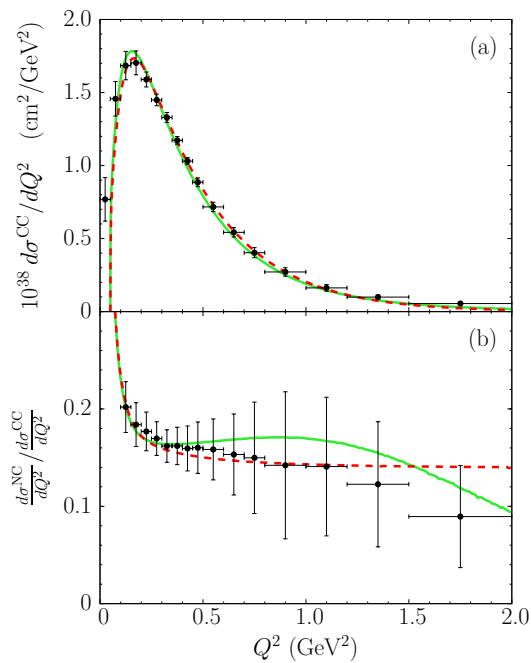}
    \subfigure{\label{fig:CC_shift_a}}
    \subfigure{\label{fig:CC_shift_b}}
\caption{\label{fig:CC_shift}(Color online) Same as Fig.~\ref{fig:CC} but for the CCQE cross section shifted by 0.05 GeV$^2$.
}
\end{figure}

Figure~\ref{fig:CC_a} shows that, as in the NCE case, our result needs to be multiplied by 1.2 to match the normalization of the \MB{} data. The agreement with the \MB{} data is fair, but the obtained cross section seems to be shifted by $-0.05\textrm{ GeV}^2$ with respect to the data, compare Figs.~\ref{fig:CC_a} and \ref{fig:CC_shift_a}. The shift size coincides with the smallest bin size of the data.

Our calculations correctly describe the ratio of the NCE to CCQE cross section in the \MB{} experiment~\cite{ref:MiniB_NC} [see Fig.~\ref{fig:CC_b}]. Most of the discrepancies are related to the shift observed in the CCQE cross section [see Fig.~\ref{fig:CC_shift_b}]. Note that the calculations using theoretical $Q^2$ cannot reproduce the shape of the NCE/CCQE ratio.

\begin{figure}
\centering
    \includegraphics[width=0.80\columnwidth]{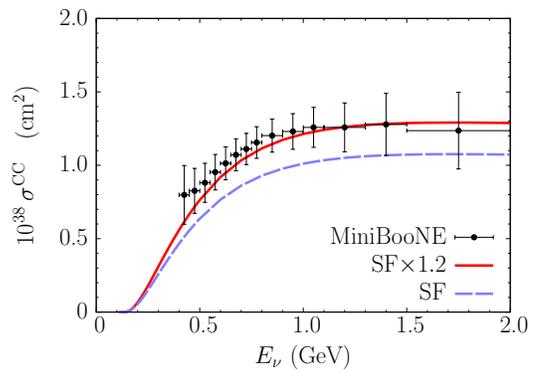}
\caption{\label{fig:tot}(Color online) The total $\isotope[12][6]{C}(\nu_\mu,\mu^-)$ cross section. Our calculations without modifications (dashed line) and multiplied by 1.2 (solid line) are compared to the \MB{} results~\cite{ref:MiniB_CC}, with the error bars showing the total uncertainty.
}
\end{figure}

For the total CCQE $\nu_\mu$ cross section, our approach yields the energy dependence in good agreement with the \MB{} result~\cite{ref:MiniB_CC}. However, as presented in Fig.~\ref{fig:tot}, also in this case the factor of 1.2 is required to match the normalization.

The contribution of the $\Delta$-production events with pion undetected, forming an irreducible background to CCQE scattering, has been subtracted in the process of extraction of the cross section and the associated uncertainties have been taken into account in the error bars of the data in Fig.~\ref{fig:tot}. For such events, the reconstructed energy is lower than the actual value, typically by $\sim$300 MeV~\cite{ref:Leitner}. One may expect their contribution to be most pronounced at the energy $\sim$450 MeV, for which the cross section receives the background from the peak of the \MB{} flux. Therefore, in Fig.~\ref{fig:tot}, the difference between the upper curve and the central value of the data for low energy is likely to be ascribable to the uncertainties related to the subtraction of the irreducible background.

From the analysis presented in this paper, it consistently emerges that, while correctly describing the shape of the \MB{}-reported cross sections (be it $d\sNC/d\qrec$, $d\sCC/d\qrec$, or $\sCC$ as a~function of neutrino energy), our calculations fail to reproduce their absolute scale, remaining inadequate by 20\%.

In this context, it is interesting to compare our results to the total $\isotope[12][6]{C}(\nu_\mu,\mu^-)$ and $\isotope[12][6]{C}(\bar\nu_\mu,\mu^+)$ cross sections measured with the recent NOMAD experiment~\cite{ref:NOMAD}. Due to the high energy of its broad-band neutrino (antineutrino) beam with the mean value of 25.9~GeV (17.6~GeV), the normalization was precisely determined from a~large sample of the deep-inelastic-scattering and inverse-muon-decay events. Using a~drift-chamber detector, a~total of 14021 (10358 single track and 3663 double track) neutrino events and 2237 antineutrino events surviving the CCQE cuts were recorded. In Monte Carlo simulations, nuclear effects were accounted for by using an approach based on the RFG model, but with realistic momentum distribution applied instead of the step function.

In the NOMAD experiment, the neutrino events with more than two tracks (the charged lepton plus one proton) detected would have been removed from the CCQE event sample.
On the other hand, in  the \MB{} experiment, all events with no pions detected, including two (or more) nucleon knockout events, contribute to the CCQE cross section.

Additional nucleons may originate from interactions between the struck nucleon and the spectator system in the final state, from the nucleon-nucleon correlations
in the initial state, and from reaction mechanisms other than nucleon knockout, such as those involving meson-exchange currents.

The effect of multiproton states due to FSI on the CCQE event sample was estimated in Monte Carlo simulations of the NOMAD experiment to be very small~\cite{ref:NOMAD}.
Because the nucleon-nucleon correlations produce in nuclei overwhelmingly proton-neutron pairs~\cite{ref:corr_science,ref:corr_Schiavilla}, they may have an influence on the track-based measurement of the CCQE cross section of neutrinos, but (in the absence of FSI) not on that of antineutrinos.

Analyzing the nucleons with momentum larger than 300~MeV,\footnote{For an event to be classified as a~double-track one in NOMAD, it was required that the proton momentum exceeds 300 MeV.} we find that in the energy range covered by the NOMAD experiment, scattering off strongly correlated nucleon pairs yields $\sim$6\% of the inclusive cross sections. This number should be considered as an upper bound of a~difference between the NOMAD- and \MB{}-reported cross sections that the approach of this paper is able to describe.

\begin{figure}
\centering
    \includegraphics[width=0.80\columnwidth]{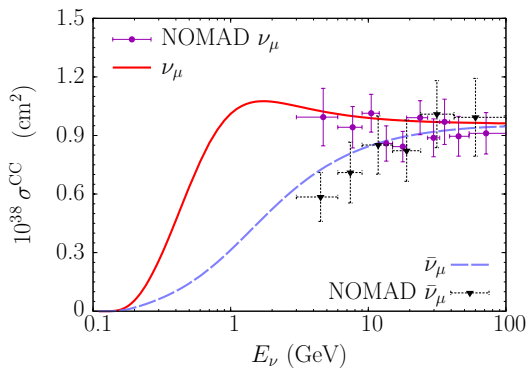}
\caption{\label{fig:NOMAD}(Color online) The total CCQE cross section for muon neutrino and antineutrino scattering off carbon. The experimental data reported by the NOMAD Collaboration are taken from Ref.~\cite{ref:NOMAD}. The error bars show the total uncertainty.
}
\end{figure}

The contribution of two- and multinucleon final states to the cross section has been thoroughly analyzed by Martini {\it et al.}~\cite{ref:Martini_CC&NC,ref:Martini_CC} and Nieves {\it et al.}~\cite{ref:Nieves_PRC,ref:Nieves_PLB}. While their results do not fully agree, both groups find that they provide significant contributions to the CCQE cross section. However, the combined analysis of the NOMAD and \MB{} data in the approaches of Refs.~\cite{ref:Martini_CC&NC,ref:Martini_CC,ref:Nieves_PRC,ref:Nieves_PLB} is not yet available.

Figure~\ref{fig:NOMAD} shows that our calculations of the total inclusive CCQE $\nu_\mu$ and $\bar\nu_\mu$ cross sections are in good agreement with the data obtained by the NOMAD Collaboration~\cite{ref:NOMAD}. We want to emphasize that their normalization is not scaled by additional factors. Although the SF results are higher by $\sim$6\% than the NOMAD best fit, this difference is less that the $\sim$8\% ($\sim$11\%) systematic uncertainty of neutrino (antineutrino) data. We observe that subtraction of the correlated contribution would bring the neutrino result in perfect agreement with the experimental points. However, this may be a~pure coincidence, as the analogical difference for antineutrinos cannot be explained in the same manner. Additionally, the results of Ref.~\cite{ref:MiniB_kappa} suggest that the approach of this paper may, to some extent, overestimate the cross section, owing to inappropriate description of the low-$Q^2$ contribution.

From the total CCQE $\nu_\mu$ cross section, the NOMAD Collaboration extracted the axial mass $1.05\pm0.02\textrm{(stat)}\pm0.06\textrm{(syst)}$ GeV. We estimate that in the SF approach that would correspond to 1.17~GeV, due to stronger quenching of the cross section. This value has been obtained without subtracting the correlated strength and, as such, gives the lowest $M_A$ required to fit the NOMAD data using our approach.

Analyzing the shape of the $\qrec$ distribution of double-track CCQE $\nu_\mu$ events for $0.2\leq \qrec\leq 4\textrm{ GeV}^2$, the NOMAD Collaboration found $M_A=1.07\pm0.06\textrm{(stat)}\pm0.07\textrm{(syst)}$ GeV. Within the quoted uncertainties, this value is consistent with our estimate based on the total cross section.

\section{Discussion}\label{sec:Discussion}
With very few exceptions~\cite{ref:BBBA07,ref:Hill}, the axial form factor $F_A(Q^2)$ is currently parametrized in the dipole form. Its value at \mbox{$Q^2=0$} is known from the neutron beta-decay measurements, whereas the dependence on $Q^2$ is governed by the axial mass $M_A$.

Because of the sizable contribution of the axial form factor to the (anti)neutrino cross section, the axial mass may be determined from the total cross sections or from shape fit to the event distribution with respect to some kinematic variable, e.g., $Q^2$. For free nucleons these two methods are equivalent, provided the dipole parametrization holds true. For bound nucleons, however, this may no longer be the case, due to inaccuracies of the applied description of nuclear effects. Note that different approaches could yield nearly identical (single) differential cross sections and, at the same time, different values of the total cross section, or {\it vice versa}.

As an illustrative example, consider CCQE (or NCE) neutrino scattering off carbon with the \MB{} beam. At $0.13\leq Q^2\leq 2.0\textrm{ GeV}^2$, the shapes of the flux-averaged cross sections $d\sCC/dQ^2$ (or $d\sNC/dQ^2$) obtained within the RFG model ($p_F=220$ MeV, $\varepsilon=34$ MeV) and the SF approach differ by less than 2.5\%, while the discrepancy between their absolute values amounts to 13\%. These features result from the fact that the $Q^2$ dependence of the cross section is governed mainly by the differential cross section on a~free nucleon [see Eq.~\eqref{eq:xsec_NC} or \eqref{eq:xsec_CC}], whereas its normalization is determined by the distribution of strength in the spectral function. Therefore, the axial mass determination from the total cross section seems to show stronger nuclear-model dependence than its extraction from the shape fit.
In addition, the procedure of flux unfolding, necessary to obtain the total cross section, introduces additional uncertainties to the result.

We made use of this observation, showing for CCQE interaction that, within the SF approach, the axial mass extracted by the \MBC{} from the first shape analysis of the $\qrec$ event distribution~\cite{ref:MiniB_kappa} is in good agreement with the total $\isotope[12][6]{C}(\nu_\mu,\mu^-)$ cross section measured with the NOMAD experiment~\cite{ref:NOMAD}.

To gauge the influence of nuclear effects on the shape of the cross section $d\sCC/d\qrec$ averaged over the \MB{} flux, we have calculated it within the SF approach and for the RFG model. In the range $0.25\leq \qrec\leq 1.0\textrm{ GeV}^2$, used in Ref.~\cite{ref:MiniB_kappa} for the determination of the axial mass within the RFG model, the SF calculation differs by less than 5\% from that for the RFG model. However, at $\qrec=2.0\textrm{ GeV}^2$ this difference increases to 12\%. Therefore, the shape of the differential cross section as a~function of $\qrec$ is affected by nuclear effects more strongly than the shape of the differential cross section as a~function of $Q^2$. We have not observed such behavior in NCE scattering, applying the definition of $\qrec$ based on knocked-out nucleon kinetic energy [see Eq.~\eqref{eq:recQ2NC}].

In Ref.~\cite{ref:MiniB_kappa}, the \MBC{} interpreted the higher value of the axial mass, extracted from the shape of the $\qrec$ distribution of CCQE(-like) $\nu_\mu$ events, as an effective method of accounting for the nuclear effects neglected in the RFG model.

Within the impulse approximation, nuclear effects tend to reduce the total cross section of muon neutrinos~\cite{ref:Davide_NPA} and slightly \emph{increase} the slope of the flux-averaged cross section $d\sCC/dQ^2$~\cite{ref:Davide_PRC}. This behavior is easy to understand: Apart from resonance excitation followed by meson absorption, the only mechanism contributing to the CCQE cross section is scattering off a~single nucleon. The nuclear targets applied in neutrino experiments are stable against the emission of nucleons; thus the nucleon separation energy is larger than the kinetic energy. For an~$\sim$1-GeV neutrino interaction, the reduction of the phase space due to the binding is therefore more pronounced than its increase resulting from Fermi motion. Moreover, the separation energy increases $Q^2$, reducing the energy transfer to the nucleon, which results in further quenching of the cross section by the nucleon form factors. For high~$Q^2$, and therefore for low-energy transfers, the effect of binding is more significant. Therefore, the interpretation of the higher value of the effective axial mass proposed by the \MBC{} cannot be supported within the IA.

However, the processes which cannot be accounted for within the IA framework have recently been shown to contribute to CCQE-like scattering, increasing the total cross section and \emph{decreasing} the slope of the $\qrec$ distribution of events.
Nieves {\it et al.}~\cite{ref:Nieves_PLB} have observed that these effects may effectively be described using a~higher value of the axial mass, confirming the interpretation of the \MBC{}.

\begin{figure}
\centering
    \includegraphics[width=0.80\columnwidth]{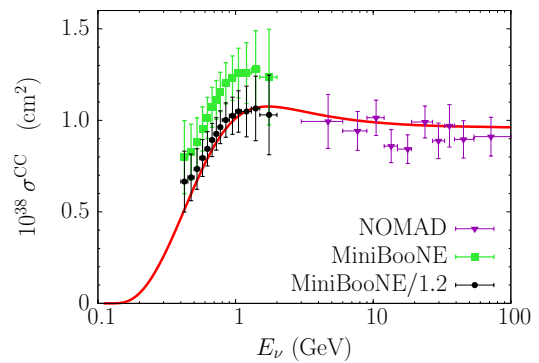}
\caption{\label{fig:tot_comb}(Color online) Comparison of the total $\isotope[12][6]{C}(\nu_\mu,\mu^-)$ cross section calculated within our approach to the data from the NOMAD~\cite{ref:NOMAD} (triangles) and \MB{}~\cite{ref:MiniB_CC} (squares) experiments. The circles show the \MB{} data divided by~1.2.
}
\end{figure}

Nevertheless, the results presented in Ref.~\cite{ref:Nieves_PLB} suggest that the neutrino flux in the \MB{} experiment is underestimated by $9.2\pm3.5$\% when $M_A=1.077$~GeV is used in the approach based on the local RFG model. Because the total cross sections obtained in the RFG model are typically higher by $\sim$10\% than those in the SF approach, the 20\% normalization discrepancy that we have observed in comparison to the \MB{} data seems to be in perfect agreement with the finding of Nieves {\it et al}~\cite{ref:Nieves_PLB}. The idea that the \MB{} flux might have been underestimated is not new~\cite{ref:Kopp}. Should it turn out to be correct, this would reconcile the \MB{} and NOMAD data in a~simple way (see Fig.~\ref{fig:tot_comb}).

Note that the \MB{} flux estimate is based on the extrapolation of the cross section for $\pi^\pm$ production in $p$-Be scattering measured on a~thin target to a~target 35 times thicker. Such calculations are known to involve severe difficulties~\cite{ref:HARP_thickTarget}.

In various interaction channels, the ratio of the events recorded with the \MB{} experiment to those predicted in the Monte Carlo simulation exceeds unity. It is interesting to note that the size of the ratio is similar: $1.21\pm0.24$ in CCQE scattering~\cite{ref:MiniB_kappa}, 1.23 in CC charged-pion production~\cite{ref:MiniB_piC}, and $1.58\pm0.05\textrm{(stat)}\pm0.26\textrm{(syst)}$ in CC neutral-pion production~\cite{ref:MiniB_pi0}. The quoted figures refer to the simulations with $M_A=1.23$~GeV and the parameter~$\kappa$, governing the enhancement of the Pauli blocking effect, set to 1.019. We checked that our approach and the model applied by \MB{} yield the total CCQE cross section differing by $\leq2.7\%$ for the neutrino energy higher than 350~MeV. Therefore, we may conclude that our 20\% normalization discrepancy is in good agreement with the \MB{} simulations and does not seem to be limited to NCE and CCQE interactions.

The latter conclusion is supported by the measurement of the inclusive CC $\nu_\mu$ cross section with the SciBooNE experiment~\cite{ref:SciBooNE}, using the same neutrino beam as \MB{}. The obtained result is larger than its Monte Carlo estimate by a~factor of 1.12 ($\kappa=1.0$) or 1.29 ($\kappa=1.022$), depending on details of the simulation.

Our calculations underestimate the absolute values of the \MB{}-reported cross sections, while being in a~good agreement with their shape and with the NOMAD data.

In the NOMAD analysis, CCQE $\nu$ events are defined as containing at most one proton detected. When the proton's kinetic energy is measured to exceed 47~MeV (momentum $\geq300$~MeV), the event is classified as a~double-track one. The cross section extracted from the single- and double-track events, composing 73.9 and 26.1\% of the collected sample, respectively, is shown not to differ. Therefore, only those two- and multinucleon final states (2NFS) which involve additional protons of kinetic energy lower than 47~MeV and any neutrons contribute to the NOMAD result. Moreover, the contributions of the 2NFS to the single- and double-track samples seem to be equal and do not show energy dependence.

In the \MB{} analysis, CCQE events are required exclusively not to involve (detected) pions. Hence, the cross sections may be increased by a~broader class of 2NFS, involving two or more protons of kinetic energy higher than 47~MeV each.

As a~consequence, the 20\% discrepancy between our calculations and the \MB{} data could, in principle, be ascribable to more sizable contributions of 2NFS than accounted for in our approach by using the effective axial mass of 1.23~GeV. Lacking apparent dependence on energy, it is, however, constrained by the NOMAD results. Note that
\begin{itemize}
\item[(i)]{the discrepancy in the NCE channel stems from the final-state nucleons with $50\leq T\leq 650$~MeV, where $T$ is their kinetic energy in total, and appears to the same extent over the whole range of $T$;}
\item[(ii)]{because the contributions of 2NFS to the \MB{}-reported NCE and CCQE cross sections seem to be equal, the kinematics of the knocked-out nucleons in these two cases should not differ significantly; and}
\item[(iii)]{the NOMAD results constrain the missing strength to involve at least two protons with $T\geq94$ MeV.}
\end{itemize}

The discrepancy between our calculations and the NCE cross section measured with \MB{} is not limited to $T\geq94$~MeV, corresponding to $\qrec\geq0.177$ GeV$^2$ and remains constant at the interval $50\leq T\leq 650$~MeV, which suggests that all the  contributing 2NFS channels are open below $T=50$~MeV. These features do not seem to be consistent with the hypothesis that 2NFS contribute to the \MB{} and NOMAD data in a~different manner, but rather point to the flux uncertainty in \MB{} being higher than reported.

\section{Summary}\label{sec:Summary}
In this paper, we have applied the spectral function approach to describe nuclear effects in (anti)neutrino scattering off carbon nucleus, treating in a~consistent manner NCE and CCQE interactions. We have considered a~broad energy range, from a~few hundreds of MeV to 100~GeV. The dipole parametrization of the axial form factor with the cutoff mass 1.23~GeV has been used, as determined from the shape of the $\qrec$ distribution of CCQE $\nu_\mu$ events by the \MBC{} in Ref.~\cite{ref:MiniB_kappa}. This effective method of accounting for two- and multinucleon final-state contributions to the cross section seems to be justified in view of recent results of Nieves {\it et al.}~\cite{ref:Nieves_PLB}.

It has been shown that our approach provides a~fairly good description of the NCE neutrino and antineutrino differential cross sections $d\sNC/dQ^2$ measured with the BNL E734 experiment. A good agreement has been found with the total CCQE $\nu_\mu$ and $\bar\nu_\mu$ cross sections reported by the NOMAD Collaboration.

Our calculations provide very accurate description of the shape of the NCE neutrino differential cross section $d\sNC/d\qrec$ obtained from the \MB{} experiment. This result confirms that nuclear effects in CCQE and NCE neutrino interactions are very similar, because the axial mass applied was extracted from the CCQE event sample. It turns out, however, that the absolute value of the calculated cross section $d\sNC/d\qrec$ underestimates the \MB{} data by 20\%. The same discrepancy is observed for CCQE $\nu_\mu$ interaction in comparisons to the differential cross section $d\sCC/d\qrec$ and the flux-unfolded total cross section reported by the \MBC{}.

The difference between the total $\isotope[12][6]{C}(\nu_\mu,\mu^-)$ cross section from NOMAD ($3\leq\Ek\leq100$~GeV) and \MB{} ($0.4\leq\Ek\leq2.0$~GeV) is sometimes attributed to a~sizable contribution of two- and multinucleon final states. This reasoning is based on the fact that in the latter experiment, nucleons knocked out from the nucleus in CCQE neutrino scattering have not been detected.

We have argued, however, that the NCE data provide an indication of the kinematics of nucleons in CCQE scattering, because the \MB{} results strongly constrain the allowed differences between nuclear effects in NCE and CCQE $\nu_\mu$ interactions. In the kinematic region used to extract the NCE cross section, we find no evidence for the contribution of two- and multinucleon final states which would not have contributed also to the CCQE cross sections reported by the NOMAD Collaboration. Therefore, the discrepancy between the results from the \MB{} and NOMAD experiments seems more likely to be ascribable to underestimated flux uncertainty in the \MB{} data analysis.

In this paper we have observed that, although the CCQE differential cross section as a~function of $Q^2$ shows very weak nuclear-model dependence, this is not the case for $d\sCC/d\qrec$. This feature may be traced back to the definition of $\qrec$ applied to CCQE interaction, as does not appear in the NCE case.

\begin{acknowledgments}
The author would like to thank Omar Benhar for illuminating discussions and critical reading of the manu{\-}script. The comments of Leslie Camilleri, Davide Meloni, and Geralyn Zeller, which helped to improve the text, are also gratefully acknowledged. This work was supported by the INFN under Grant No. MB31.
\end{acknowledgments}

\end{document}